\documentclass[runningheads]{llncs}

\def\myauthors{David Otero, Javier Parapar}
\def\mytitle{LLM-Assisted Pseudo-Relevance Feedback}

\usepackage{multirow}
\usepackage{soul}
\usepackage{subcaption}
\usepackage{amsmath}
\usepackage{graphicx}
\usepackage{url}
\usepackage[inline]{enumitem}
\usepackage{acronym}
\usepackage{booktabs}
\usepackage{longtable}
\usepackage{pdflscape}
\usepackage{rotating}
\usepackage{float}
\usepackage{tabularx}
\usepackage{afterpage}
\usepackage{siunitx}
\usepackage{algpseudocode, algorithm}
\usepackage{calc}
\usepackage[hidelinks]{hyperref}
\usepackage{todonotes}
\usepackage{datenumber}
\usepackage[space,sort]{cite}
\usepackage[multiple]{footmisc}
\usepackage[T1]{fontenc}
\usepackage{cleveref}
\usepackage{fontawesome}

\author{David Otero\orcidID{0000-0003-1139-0449} \and Javier Parapar\orcidID{0000-0002-5997-8252}}

\institute{IRLab, CITIC, Universidade da Coruña, Spain\\ \email{david.otero.freijeiro@udc.es} \email{javier.parapar@udc.es}}

\authorrunning{D. Otero and J. Parapar}

\hypersetup{
  pdfauthor={\myauthors},
  pdftitle={\mytitle},
  pdfsubject={Information Retrieval},
}

\graphicspath{{plots/}}
\newcommand{\ra}[1]{\renewcommand{\arraystretch}{#1}}

\newcommand{\myparagraph}[1]{\paragraph*{\normalsize\bf#1}}

\algblock{Input}{EndInput} % Create new block with start and end command
\algnotext{EndInput}  % Hide end input command
\algblock{Output}{EndOutput} % Create new block with start and end command
\algnotext{EndOutput} % Hide end output command

\newsavebox\CBox
\def\textBF#1{\sbox\CBox{#1}\resizebox{\wd\CBox}{\ht\CBox}{\textbf{#1}}}

\begin{document}

%%%%%%%%%%%%%%%%%%%%%%%%%%%%%%%%%%%%%%%%%%%%%%%%%
% Title
%%%%%%%%%%%%%%%%%%%%%%%%%%%%%%%%%%%%%%%%%%%%%%%%%
\title{\mytitle}
\titlerunning{LLM-Assisted Pseudo-Relevance Feedback}

\maketitle

%%%%%%%%%%%%%%%%%%%%%%%%%%%%%%%%%%%%%%%%%%%%%%%%%
% Abstract
%%%%%%%%%%%%%%%%%%%%%%%%%%%%%%%%%%%%%%%%%%%%%%%%%
\begin{abstract}
Query expansion is a long-standing technique to mitigate vocabulary mismatch in ad hoc Information Retrieval. Pseudo-relevance feedback methods, such as RM3, estimate an expanded query model from the top-ranked documents, but remain vulnerable to topic drift when early results include noisy or tangential content. Recent approaches instead prompt Large Language Models to generate synthetic expansions or query variants. While effective, these methods risk hallucinations and misalignment with collection-specific terminology. We propose a hybrid alternative that preserves the robustness and interpretability of classical PRF while leveraging LLM semantic judgement. Our method inserts an LLM-based filtering stage prior to RM3 estimation: the LLM judges the documents in the initial top-$k$ ranking, and RM3 is computed only over those accepted as relevant. This simple intervention improves over blind PRF and a strong baseline across several datasets and metrics.

\end{abstract}

%%%%%%%%%%%%%%%%%%%%%%%%%%%%%%%%%%%%%%%%%%%%%%%%%
% Sections
%%%%%%%%%%%%%%%%%%%%%%%%%%%%%%%%%%%%%%%%%%%%%%%%%
% !TeX spellcheck = en_GB

%%%%%%%%%%%%%%%%%%%%%%%%%%%%%%%%%%%%%%%%%%%%%%%%%
% Introduction
%%%%%%%%%%%%%%%%%%%%%%%%%%%%%%%%%%%%%%%%%%%%%%%%%
\section{Introduction}
\label{sec:intro}

Information Retrieval (IR) systems continue to struggle with vocabulary mismatch, a problem arising when users express their queries with keywords that do not match the terms used in relevant documents. Query expansion (QE) tries to address this problem by enriching the user query with terms that better capture the underlying intent~\cite{Carpineto2012}. Pseudo-relevance feedback (PRF) is a QE technique that assumes the top-$k$ documents from a first retrieval pass are relevant, and exploits this information to come up with an expanded query. Among probabilistic PRF approaches, relevance-based language modelling~\cite{Lavrenko2001} (e.g., RM3~\cite{NasreenJaleel2004}) remains a strong, competitive baseline~\cite{Lv2009a}. Specifically, RM3 extracts probabilistic term evidence from the assumed-relevant documents and interpolates it with the original query to form an expanded query, which is re-issued to the search system to produce the final ranking. PRF is an studied approach that may work generally well, but noisy documents in the top-$k$ can casuse topic drift. Topic drift refers to the situation where the expanded query no longer represents the original intent of the user, this is, it has \textit{drifted} away.

Recent work has revisited QE by using Large Language Models (LLMs)~\cite{Lei2024,Alaofi2025,LiangWang2023,Jagerman2023,XiaoWang2023,Alaofi2023}. Instead of employing corpus evidence, LLM-based prompting techniques generate hypothetical documents, reformulations, or query variants~\cite{Jagerman2023,LiangWang2023,XiaoWang2023}. These advances can substantially improve effectiveness, but they introduce critical risks: LLM-generated content is prone to hallucinations, producing terms or concepts that do not exist in the target collection. This misalignment with collection-specific terminology can increase the topic drift, where generated text introduces useless information~\cite{Abe2025}. While hybrid strategies seek to constrain LLM output with corpus signals~\cite{Lei2024}, the fundamental issue remains that generative approaches rely on the model producing grounded, relevant content rather than leveraging the actual corpus evidence.

In this work, we explore a different hybrid strategy: instead of using LLMs to generate expansion terms, we exploit their strong judging capabilities to filter the pseudo-relevant set before applying RM3. Specifically, we insert an LLM-based classifier as a filter over the initially retrieved candidate documents. The model judges each document's relevance to the query, and only documents classified as relevant are included in the feedback set for estimation. This design isolates the powerful semantic discrimination capabilities of LLMs to the task of denoising feedback evidence, while preserving the robustness, interpretability, and established effectiveness of classical probabilistic expansion.

We evaluate\footnote{Code available at \faicon{github} \href{https://github.com/davidoterof/ecir2026}{https://github.com/davidoterof/ecir2026}} on multiple datasets of different nature and observe consistent improvements of our proposals over strong baselines. Our method offers a simple alternative that leverages LLM judgment while remaining grounded in corpus evidence, largely avoiding hallucination issues typical of generative approaches.
% !TeX spellcheck = en_GB

%%%%%%%%%%%%%%%%%%%%%%%%%%%%%%%%%%%%%%%%%%%%%%%%%
% Method
%%%%%%%%%%%%%%%%%%%%%%%%%%%%%%%%%%%%%%%%%%%%%%%%%
\section{Method}
\label{sec:method}

Pseudo-relevance feedback assumes the top-$k$ documents from a first retrieval pass are relevant, and uses them to extract expansion terms to improve the query. We adopt RM3. Given the set of candidate documents $D_k$, RM3 estimates a relevance model $P(t|R)$ from $D_k$ and interpolates it with the original query model $P(t|q)$ to obtain the expanded query model $P_{RM3}(t)$. The expanded query weights are computed as:

\begin{equation}
\label{eq:interpolation}
  P_{RM3}(t) = \lambda P(t|q) + (1 - \lambda) P(t|R)
\end{equation}

\noindent where $\lambda \in [0,1]$ is an interpolation parameter. To estimate the query model $P(t|q)$ we only use the query text. The relevance model $P(t|R)$ is:

\begin{equation}
  P(t|R) \propto \sum_{d\in D_k} P(d) \cdot P(t|d) \cdot \prod_{w\in q} P(w|d),
\end{equation}

\noindent  where $P(d)$ is a (usually uniform) document prior, and thus we ignore it for ranking, $P(t|d)$ is the document language model, and $P(w|d)$ the query likelihood term probability. In practice, we keep the top-$e$ terms from $P_{RM3}(t)$ after normalisation.

After the firs retrieval pass, the original query is expanded and reweighted according to \Cref{eq:interpolation} and reissued to the search system.

\subsection{Method 1: LLM-based Filtering}

Early ranked documents may include noise, diffusing $P(t|R)$, and harming the results of the second pass. Our first contribution consists in using a model to filter the documents in $D_k$.  We insert a lightweight LLM-based filtering stage between initial retrieval and RM3 estimation to denoise $D_k$:

\begin{enumerate}
  \item Retrieve an initial ranked list with a lexical baseline (e.g., query likelihood with Dirichlet).
  \item Take the top-$k$ documents as candidate pseudo-relevant documents $D_k$.
  \item For each $d \in D_k$, obtain an LLM binary judgement $y_d\in\{0,1\}$ from the query–document pair.
  \item Construct a filtered set $D_k^{\mathrm{LLM}}=\{d\in D_k: y_d=1\}$.
  \item Apply the standard RM3 estimation steps, replacing $D_k$ with $D_k^{LLM}$, compute $P(t|R)$ over the filtered set and interpolate with the original query model and rerank.
\end{enumerate}

The key advantage of this approach is that LLM judgement capabilities improve the quality of the pseudo-relevance assumption without introducing external content. This differs from LLM-generative methods in several ways. First, we do not generate any text, greatly reducing the risk of hallucinations. Second, all expansion terms come from actual corpus documents judged as relevant by the model, ensuring grounding in collection-specific terminology.

\subsection{Method 2: LLM-filtered and Probability-weighted PRF}

We develop a different variant of our Method 1 that builds upon it. This variant first applies Method 1 to build $D_k^{\mathrm{LLM}}$ and then, at expansion/ranking time, plugs the token probability that the LLM assigned to the token \textit{`true'}.\footnote{We use the same prompt in Method 1 and Method 2. The difference is that in the first method we use the relevance label and in the second method we use the probability that the LLM gave to the token \textit{`true'}.} Starting from Eq.~(2), which uses the query-likelihood product $\prod_{w\in q} P(w\mid d)$, we replace that factor by $P_{LLM}(\text{next token}=\texttt{`true'}\mid q,d)$:
\begin{equation*}
  P(t\mid R) \propto \sum_{d\in D_k^{\mathrm{LLM}}} P(t\mid d) \cdot P_{LLM}(\text{next token}=\texttt{`true'}\mid q,d),
\end{equation*}
Eq.~(1) for interpolation remains unchanged.

%%%%%%%%%%%%%%%%%%%%%%%%%%%%%%%%%%%%%%%%%%%%%%%%%
% Experiments
%%%%%%%%%%%%%%%%%%%%%%%%%%%%%%%%%%%%%%%%%%%%%%%%%
\section{Experiments}
\label{sec:experiments}

In this section, we explain the details of the experiments we carry out to evaluate the performance of our proposed methods.

\subsection{Experimental Setup}

\myparagraph{Collections.}

We consider several standard IR test collections: a subset of the Associated Press collection corresponding to the 1988 and 1989 years, the ROBUST04 collection~\cite{Voorhees2004}, and the MSMARCO dataset~\cite{Bajaj2016}, both with the DL-19~\cite{Craswell2019} and DL-20~\cite{Craswell2020} NIST topics. We have summarised this information in \Cref{tab:datasets}.

\begin{table}[t]
  \centering
  \small
  \setlength{\tabcolsep}{4pt}
  \caption{Collections and topics used for experimentation.}
  \label{tab:datasets}
  \begin{tabular}{lrr}
    \toprule
    {Dataset} & {Training Topics} & {Test Topics} \\
    \midrule
    AP88-89 & 51-100 & 101-200 \\
    ROBUST04 & 301-350 & 351-400 \\
    MSMARCO & DL-19  & DL-20 \\
    WT10G & 451-500 & 501-550 \\
    \bottomrule
  \end{tabular}
\end{table}

\myparagraph{Compared methods.}

We compare the following methods:
(i) Sparse retrieval with Dirichlet priors (QLD)~\cite{Ponte1998};
(ii) RM3 over the unfiltered top-$k$ (RM3);
(iii) MonoT5 re-ranking (MonoT5 rerank);
(iv) Filter the PRF set with MonoT5 labels (MonoT5F);
(v) MonoT5F plus plugging the next-token probability into query likelihood for RM3 estimation (MonoT5F + RM3 w/prob);
(vi) The same two variants using Llama 3.1-8B-Instruct (LLMF, LLMF + RM3 w/prob).

\myparagraph{Evaluation.}

\sloppy We report Average Precision (AP) and normalized discounted cumulative gain (NDCG). We perform statistical testing using the Wilcoxon Signed-Rank text with Benjamini-Hochberg correction for multiple comparisons\cite{Otero2025}.

\myparagraph{Tranining and Evaluation.}

We follow a training and test strategy on different topics splits. The details of which splits we use in each collection are show in \Cref{tab:datasets}. The training topics are used to tune the hyperparameters of the different methods, optimizing for AP@1000. The parameters we tune are the following: the number of documents used for PRF $k \in \{100, 75, 50, 25, 10, 5\}$; the number of terms selected for expansion $e \in \{5, 10, 15, 20, 25, 30\}$; and the interpolation parameter $\lambda \in \{0.1, 0.2, 0.3, 0.4, 0.5, 0.6, 0.7, 0.8, 0.9\}$. These parameters were tuned separately for each method and dataset.

\subsection{Results}

\Cref{tab:results} reports AP@1000 and NDCG@100 for the four collections across eight methods (the presented baselines and proposals, plus an oracle upper bound). The oracle method is included to represent the upper bound performance we can obtain with RM3. This oracle takes the true relevant documents from $D_k$ and then applies RM3. We also tuned the PRF parameters of this method. We summarise the results with the following highlights:
(i) RM3 improves over QLD on all datasets, but less so on DL-20 where early precision drops slightly;
(ii) MonoT5 re-ranking yields the best effectiveness on DL-20, especially on NDCG@100, but lags behind on the other collection; DL20 is the only collection where MonoT5 is the top method, highlighting the strong domain dependence on the MSMARCO training data;
(iii) filtering the PRF set with MonoT5 consistently lifts both AP and NDCG over RM3, with the largest gains on ROBUST04 and WT10G;
(iv) adding the next token probability to RM3 further improves AP and NDCG on AP88–89 and WT10G, while remaining competitive on ROBUST04;
(v) using a generic LLM to filter (LLMF) also helps over RM3, but lags behind MonoT5F; this is expected, since MonoT5 was fine-tuned for the relevance task, whereas the generic LLM was not.
(vi) the oracle shows substantial headroom, suggesting further improvements are possible with better filtering.

\begin{table}
\centering
\small
\setlength{\tabcolsep}{2pt}
\ra{1.1}
\caption{Effectiveness results with AP@1000 and NDCG@100. Best model per column is \textbf{boldfaced}. Statistically significant improvements are indicated with the corresponding superscript above the better model.}
\label{tab:results}
\resizebox{\columnwidth}{!}{%
\begin{tabular}{@{}lrrrrrrrr@{}}
  \toprule
  \multirow{2.3}{*}{Method} & \multicolumn{4}{c}{AP@1000} & \multicolumn{4}{c}{NDCG@100} \\
  \cmidrule(lr){2-5} \cmidrule(lr){6-9}
  \addtolength{\tabcolsep}{-1pt}
     & \small AP8889 & \small R04 & \small WT10G & \small DL-20 & \small AP8889 & \small R04 & \small WT10G & \small DL-20 \\
  \midrule
    $(a)$ QLD                                      & 0.2213 & 0.1829 & 0.1519 & 0.3287 & 0.3774 & 0.3314 & 0.3041 & 0.4781 \\
                                                   & \scriptsize \textit{a,c} & \scriptsize \textit{a} & \scriptsize \textit{a} &  & \scriptsize \textit{a,c} & \scriptsize \textit{} & \scriptsize \textit{a} &  \\
    $(b)$ ~ + RM3                                  & 0.2920 & 0.2004 & 0.1587 & 0.3369 & 0.4419 & 0.3446 & 0.3166 & 0.4642 \\
                                                   & \scriptsize \textit{a} &  & \scriptsize \textit{a} & \scriptsize \textit{a,b,d,f,g} & \scriptsize \textit{a} &  & \scriptsize \textit{a} & \scriptsize \textit{a,b,f,g} \\
    $(c)$ ~ + MonoT5 rerank                        & 0.2474 & 0.1907 & 0.1617 & \textBF{0.4256} & 0.4016 & 0.3356 & 0.3174 & \textBF{0.5622} \\
                                                   & \scriptsize \textit{a,c} & \scriptsize \textit{a,b,c,f,g} & \scriptsize \textit{a,b,c,f,g} & \scriptsize \textit{a,b,f} & \scriptsize \textit{a,c} & \scriptsize \textit{a,b,c,f,g} & \scriptsize \textit{a,b,c,f,g} & \scriptsize \textit{abf} \\
    $(d)$ ~ + MonoT5F + RM3                        & 0.2975 & \textBF{0.2317} & 0.2042 & 0.3941 & 0.4503 & \textBF{0.3888} & 0.3651 & 0.5317 \\
                                                   & \scriptsize \textit{a,b,c} & \scriptsize \textit{a,c} & \scriptsize \textit{a,b,c,f,g} & \scriptsize \textit{a,b,d,f,g} & \scriptsize \textit{a,c} & \scriptsize \textit{a,b,c} & \scriptsize \textit{a,b,c,f,g} & \scriptsize \textit{a,b,f,g} \\
    $(e)$ ~ + MonoT5F + RM3 w/prob                 & \textBF{0.3072} & 0.2214 & \textBF{0.2067} & 0.4154 & \textBF{0.4612} & 0.3717 & \textBF{0.3680} & 0.5355 \\
                                                   & \scriptsize \textit{a,c} & \scriptsize \textit{a,b,g} & \scriptsize \textit{a,b} &  & \scriptsize \textit{a,c} & \scriptsize \textit{a,b,c} & \scriptsize \textit{a} &  \\
    $(f)$ ~ + LLMF + RM3                           & 0.3002 & 0.2137 & 0.1720 & 0.3722 & 0.4515 & 0.3644 & 0.3254 & 0.4913 \\
                                                   & \scriptsize \textit{a,c} & \scriptsize \textit{a,b} & \scriptsize \textit{a,b} & \scriptsize \textit{a} & \scriptsize \textit{a,c} & \scriptsize \textit{a,b} &  &  \\
    $(g)$ ~ + LLMF + RM3 w/prob                    & 0.2962 & 0.2091 & 0.1747 & 0.3683 & 0.4493 & 0.3587 & 0.3278 & 0.4931 \\

    \midrule
    ~~~~~~ + RM3 ORACLE                        & 0.3845 & 0.2747 & 0.2831 & 0.4350 & 0.5669 & 0.4535 & 0.4634 & 0.5696 \\

  \bottomrule
\end{tabular}%
}
\end{table}

\myparagraph{Prompt ablation with Llama.}
In the experiments above, we used only title queries to prompt the LLM. However, TREC title queries are typically short, and relying solely on the title can make relevance ambiguous. The TREC topic narrative, by contrast, provides assessors with explicit guidance on what should be considered relevant. In this section, we investigate the effect of incorporating the narrative into the prompt. Specifically, we augment the judgment prompt with the narrative, providing Llama with the title, the narrative (as relevance instructions), and the candidate document text. All other experimental settings remain identical to those in the previous section. Results on the test topics are reported in \Cref{tab:llama-prompt}. Since DL-20 lacks narratives, it is excluded from this analysis. Note that RM3 was not provided with this additional information, so its query input remains the short (title-only) version. In other words, the narrative was used exclusively in the prompt issued to Llama for judging. Overall, including the narrative consistently improves the LLM-filtered variants over their title-only counterparts on AP88–89 and ROBUST04 for both AP@1000 and NDCG@100, with the largest gains observed when weighting RM3 with the next token probability. On WT10G, the narrative also proves beneficial, particularly for early precision (NDCG@100). With this change, the LLM-based methods that use the narrative become the top-performing approaches across nearly all collections and metrics, achieving statistically significant improvements over most competing methods. As a closer look, we also compute the Robustness Index, which measures the number of queries both improved and degraded with respect to the baseline. We do not report it here because of space constraints. We observed that including the narrative section in the prompt of the LLM resulted in a severe drop in the number of damaged queries.

\begin{table}
\centering
\small
\setlength{\tabcolsep}{2pt}
\ra{1.1}
\caption{Effectiveness results obtained with LLMs \textbf{when including the narrative section in the prompt}. Percentage indicates the improvement over the title-only alternative. DL-20 results are not reported as narratives are not available. Best model per column is boldfaced. Statistically significant improvements are indicated with the corresponding superscript above the better model.}
\label{tab:llama-prompt}
\resizebox{\columnwidth}{!}{%
\begin{tabular}{@{}lcccccc@{}}
  \toprule
  \multirow{2.3}{*}{Method} & \multicolumn{3}{c}{AP@1000} & \multicolumn{3}{c}{NDCG@100} \\
  \cmidrule(lr){2-4} \cmidrule(lr){5-7}
  \addtolength{\tabcolsep}{-1pt}
  & \small AP8889 & \small R04 & \small WT10G & \small AP8889 & \small R04 & \small WT10G \\
  \midrule
                                                   & \scriptsize \textit{a,b,c,d,e,f,g} & \scriptsize \textit{a,b,c,f,g} & \scriptsize \textit{a,b,c,f,g} & \scriptsize \textit{a,b,c,d,e,f,g} & \scriptsize \textit{a,b,c,f,g} & \scriptsize \textit{a,b,c,f,g}  \\
    $(h)$ ~ + LLMF + RM3                           & 0.3338 {\scriptsize$\uparrow11\%$} & 0.2390 {\scriptsize$\uparrow12\%$}& \textBF{0.2115} {\scriptsize$\uparrow23\%$}& 0.4949 {\scriptsize$\uparrow10\%$} & 0.3992 {\scriptsize$\uparrow10\%$}& \textBF{0.3890} {\scriptsize$\uparrow20\%$}\\
                                                   & \scriptsize \textit{a,b,c,d,e,f,g,h} & \scriptsize \textit{a,b,c,f,g} & \scriptsize \textit{a,b,c,f,g} & \scriptsize \textit{a,b,c,d,e,f,g,h} & \scriptsize \textit{a,b,c,e,f,g} &  \scriptsize \textit{a,b,c,f,g} \\
    $(i)$ ~ + LLMF + RM3 w/prob                    & \textBF{0.3540} {\scriptsize$\uparrow20\%$}& \textBF{0.2441} {\scriptsize$\uparrow17\%$}&  0.2056 {\scriptsize$\uparrow18\%$}& \textBF{0.5125} {\scriptsize$\uparrow14\%$}& \textBF{0.4033} {\scriptsize$\uparrow12\%$}&  0.3791 {\scriptsize$\uparrow16\%$}\\
  \bottomrule
\end{tabular}%
}
\end{table}

% !TeX spellcheck = en_GB

%%%%%%%%%%%%%%%%%%%%%%%%%%%%%%%%%%%%%%%%%%%%%%%%%
% Conclusions
%%%%%%%%%%%%%%%%%%%%%%%%%%%%%%%%%%%%%%%%%%%%%%%%%
\section{Conclusions}
\label{sec:conclusions}

We introduced an LLM-assisted filtering stage for pseudo-relevance feedback that preserves the simplicity of the classical RM3 estimator while addressing the topic-drift that may affect to LLM-generation based methods. By restricting RM3 estimation to LLM-accepted feedback documents, the expansion vocabulary remains corpus-grounded yet less polluted by tangential or popular but irrelevant terms. Empirical results across all evaluated collections show that the proposed LLM-filtered RM3 consistently and significantly outperforms standard RM3 and a strong reranking model in both AP and NDCG. While the introduction of an LLM-based filtering stage may incur a moderate computational overhead, this cost is limited to a small subset of top-ranked documents and remains compatible with practical retrieval pipelines. We believe that a more detailed analysis of this overhead and its trade-offs with effectiveness is an interesting direction for future work. Our experiments also showed that including the narrative section of TREC topics in the LLM prompt improved the overall performance of LLM-based filtering approaches, with a great reduction in the number of damaged queries with respect to the baseline. Our findings suggest that selectively integrating LLM reasoning as a discriminative filter is a promising direction for enhancing classical IR models without abandoning their interpretability and efficiency. We believe that several lines are now open for future work. The first one is performing a more detailed analysis of the effects of the PRF parameters in othe overall performance. We would also like to explore the use of stronger LLM models and reasoning-powered alternatives.

%%%%%%%%%%%%%%%%%%%%%%%%%%%%%%%%%%%%%%%%%%%%%%%%%
% Credits
%%%%%%%%%%%%%%%%%%%%%%%%%%%%%%%%%%%%%%%%%%%%%%%%%
\begin{credits}
\subsubsection{\ackname}

The authors thank the financial support supplied by the grant PID2022-137061OB-C21 funded by MI-CIU/AEI/10.13039/501100011033 and by “ERDF/EU”. The authors also thank the funding supplied by the Consellería de Cultura, Educación, Formación Profesional e Universidades (accreditations ED431G 2023/01 and ED431C 2025/49) and the European Regional Development Fund, which acknowledges the CITIC, as a center accredited for excellence within the Galician University System and a member of the CIGUS Network, receives subsidies from the Department of Education, Science, Universities, and Vocational Training of the Xunta de Galicia. Additionally, it is co-financed by the EU through the FEDER Galicia 2021-27 operational program (Ref. ED431G 2023/01).

\subsubsection{\discintname}
The authors have no competing interests to declare that are relevant to the content of this article.
\end{credits}

%%%%%%%%%%%%%%%%%%%%%%%%%%%%%%%%%%%%%%%%%%%%%%%%%
% References
%%%%%%%%%%%%%%%%%%%%%%%%%%%%%%%%%%%%%%%%%%%%%%%%%
\bibliography{references}

@inproceedings{Ponte1998,
  booktitle = {Proceedings of the 21st Annual International ACM SIGIR Conference on Research and Development in Information Retrieval},
  author = {Ponte, Jay M. and Croft, W. Bruce},
  title = {A Language Modeling Approach to Information Retrieval},
  year = {1998},
  pages = {275--281},
  numpages = {7},
  publisher = {Association for Computing Machinery},
  address = {New York, NY, USA},
  location = {Melbourne, Australia},
  series = {SIGIR '98},
  isbn = {1581130155},
  doi = {10.1145/290941.291008},
}

@inproceedings{Lavrenko2001,
  booktitle = {Proceedings of the 24th Annual International ACM SIGIR Conference on Research and Development in Information Retrieval},
  author = {Lavrenko, Victor and Croft, W. Bruce},
  title = {Relevance Based Language Models},
  year = {2001},
  pages = {120--127},
  numpages = {8},
  publisher = {Association for Computing Machinery},
  address = {New York, NY, USA},
  location = {New Orleans, Louisiana, USA},
  series = {SIGIR '01},
  isbn = {1581133316},
  doi = {10.1145/383952.383972},
}

@inproceedings{NasreenJaleel2004,
  booktitle = {Proceedings of TREC 2004},
  title = {{UMass} at {TREC} 2004: Novelty and {HARD}},
  author = {Abdul-Jaleel, Nasreen and Allan, James and Croft, W. Bruce and Diaz, Fernando and Larkey, Leah and Li, Xiaoyan and Smucker, Mark D. and Wade, Courtney},
  year = {2004},
  pages = {1--13},
  url = {https://trec.nist.gov/pubs/trec13/papers/umass.novelty.hard.pdf}
}

@inproceedings{Lv2009a,
  booktitle = {Proceedings of the 18th ACM Conference on Information and Knowledge Management},
  author = {Lv, Yuanhua and Zhai, ChengXiang},
  title = {A Comparative Study of Methods for Estimating Query Language Models with Pseudo Feedback},
  year = {2009},
  pages = {1895--1898},
  numpages = {4},
  publisher = {Association for Computing Machinery},
  address = {New York, NY, USA},
  location = {Hong Kong, China},
  series = {CIKM '09},
  isbn = {9781605585123},
  doi = {10.1145/1645953.1646259},
}

@inproceedings{Craswell2019,
  booktitle  = {Proceedings of the Twenty-Eight Text REtrieval Conference},
  author     = {Craswell, Nick and Mitra, Bhaskar and Yilmaz, Emine and Campos, Daniel and Voorhees, Ellen M.},
  title      = {Overview of the {TREC} 2019 Deep Learning Track},
  year       = {2019},
  publisher  = {NIST Special Publication 1250},
  address    = {Gaithersburg, Maryland, USA},
  url        = {https://trec.nist.gov/pubs/trec28/papers/OVERVIEW.DL.pdf}
}

@inproceedings{Craswell2020,
  booktitle  = {Proceedings of the Twenty-Ninth Text REtrieval Conference},
  author     = {Craswell, Nick and Mitra, Bhaskar and Yilmaz, Emine and Campos, Daniel},
  title      = {Overview of the {TREC} 2020 Deep Learning Track},
  year       = {2020},
  publisher  = {NIST Special Publication 1266},
  address    = {Gaithersburg, Maryland, USA},
  url        = {https://trec.nist.gov/pubs/trec29/papers/OVERVIEW.DL.pdf}
}

@article{Carpineto2012,
  author = {Carpineto, Claudio and Romano, Giovanni},
  title = {A Survey of Automatic Query Expansion in Information Retrieval},
  year = {2012},
  publisher = {Association for Computing Machinery},
  address = {New York, NY, USA},
  volume = {44},
  number = {1},
  issn = {0360-0300},
  doi = {10.1145/2071389.2071390},
  journal = {ACM Computing Surveys},
  numpages = {50},
}

@inproceedings{Alaofi2023,
  author = {Alaofi, Marwah and Gallagher, Luke and Sanderson, Mark and Scholer, Falk and Thomas, Paul},
  title = {Can Generative LLMs Create Query Variants for Test Collections? An Exploratory Study},
  booktitle = {Proceedings of the 46th International ACM SIGIR Conference on Research and Development in Information Retrieval},
  year = {2023},
  isbn = {9781450394086},
  publisher = {Association for Computing Machinery},
  address = {New York, NY, USA},
  doi = {10.1145/3539618.3591960},
  pages = {1869--1873},
  numpages = {5},
  location = {Taipei, Taiwan},
  series = {SIGIR '23}
}

@inproceedings{LiangWang2023,
  title = {Query2doc: Query Expansion with Large Language Models},
  author = {Wang, Liang  and Yang, Nan  and Wei, Furu},
  booktitle = {Proceedings of the 2023 Conference on Empirical Methods in Natural Language Processing},
  year = {2023},
  publisher = {Association for Computational Linguistics},
  doi = {10.18653/v1/2023.emnlp-main.585},
  pages = {9414--9423},
}

@inproceedings{Jagerman2023,
  author    = {Jagerman, Rolf and Zhuang, Honglei and Qin, Zhen and Wang, Xuanhui and Bendersky, Michael},
  title     = {Query Expansion by Prompting Large Language Models},
  year      = {2023},
  maintitle = {Proceedings of the 46th International ACM SIGIR Conference on Research and Development in Information Retrieval},
  booktitle = {The First Workshop on Generative Information Retrieval (GenIR@SIGIR 2023)},
}

@inproceedings{XiaoWang2023,
  author = {Wang, Xiao and MacAvaney, Sean and Macdonald, Craig and Ounis, Iadh},
  title = {Generative Query Reformulation for Effective Adhoc Search},
  year      = {2023},
  maintitle = {Proceedings of the 46th International ACM SIGIR Conference on Research and Development in Information Retrieval},
  booktitle = {The First Workshop on Generative Information Retrieval (GenIR@SIGIR 2023)},
}

@inproceedings{Otero2025,
  title={Towards Reliable Testing for Multiple Information Retrieval System Comparisons},
  author={Otero, David and Parapar, Javier and Barreiro, {\'A}lvaro},
  year={2025},
  booktitle={Advances in Information Retrieval},
  publisher={Springer Nature Switzerland},
  address={Cham},
  pages={424--439},
  isbn={978-3-031-88711-6}
}

@inproceedings{Voorhees2004,
  author = {Voorhees, Ellen M.},
  title = {Overview of the {TREC} 2004 Robust Track},
  booktitle = {Proceedings of the Thirteenth Text REtrieval Conference, {TREC} 2004, Gaithersburg, Maryland, USA, November 16-19, 2004},
  series = {{NIST} Special Publication},
  volume = {500-261},
  publisher = {National Institute of Standards and Technology {(NIST)}},
  year = {2004},
  url = {http://trec.nist.gov/pubs/trec13/papers/ROBUST.OVERVIEW.pdf},
}

@inproceedings{Abe2025,
  author = {Abe, Kenya and Takeoka, Kunihiro and Kato, Makoto P. and Oyamada, Masafumi},
  title = {LLM-based Query Expansion Fails for Unfamiliar and Ambiguous Queries},
  year = {2025},
  isbn = {9798400715921},
  publisher = {Association for Computing Machinery},
  address = {New York, NY, USA},
  doi = {10.1145/3726302.3730222},
  booktitle = {Proceedings of the 48th International ACM SIGIR Conference on Research and Development in Information Retrieval},
  pages = {3035--3039},
  numpages = {5},
  location = {Padua, Italy},
  series = {SIGIR '25}
}

@inproceedings{Lei2024,
    title = {Corpus-Steered Query Expansion with Large Language Models},
    author = {Lei, Yibin  and Cao, Yu  and Zhou, Tianyi  and Shen, Tao  and Yates, Andrew},
    booktitle = {Proceedings of the 18th Conference of the European Chapter of the Association for Computational Linguistics (Volume 2: Short Papers)},
    year = {2024},
    publisher = {Association for Computational Linguistics},
    doi = {10.18653/v1/2024.eacl-short.34},
    pages = {393--401},
}

@article{Bajaj2016,
  title={Ms marco: A human generated machine reading comprehension dataset},
  author={Bajaj, Payal and Campos, Daniel and Craswell, Nick and Deng, Li and Gao, Jianfeng and Liu, Xiaodong and Majumder, Rangan and McNamara, Andrew and Mitra, Bhaskar and Nguyen, Tri and others},
  journal={arXiv preprint arXiv:1611.09268},
  year={2016}
}

@inproceedings{Alaofi2025,
    author = {Alaofi, Marwah and Ferro, Nicola and Thomas, Paul and Scholer, Falk and Sanderson, Mark},
    title = {Demographically-Inspired Query Variants Using an LLM},
    year = {2025},
    isbn = {9798400718618},
    publisher = {Association for Computing Machinery},
    address = {New York, NY, USA},
    doi = {10.1145/3731120.3744608},
    booktitle = {Proceedings of the 2025 International ACM SIGIR Conference on Innovative Concepts and Theories in Information Retrieval (ICTIR)},
    pages = {390--400},
    numpages = {11},
    location = {Padua, Italy},
    series = {ICTIR '25}
}
\bibliographystyle{splncs04}

\end{document}